\documentclass[12pt]{article}

\usepackage{amsfonts}
\usepackage{graphicx}
\usepackage{amssymb}
\usepackage{amsmath}
\usepackage{epsfig}
\usepackage[usenames]{color}

\newcommand{\be}{\begin{equation}}
\newcommand{\ee}{\end{equation}}
\newcommand{\bea}{\begin{eqnarray}}
\newcommand{\eea}{\end{eqnarray}}
\newcommand{\ba}{\begin{array}}
\newcommand{\nno}{\nonumber}
\newcommand{\ea}{\end{array}}
\newcommand{\eq}[1]{(\ref{#1})}

\numberwithin{equation}{section}

\begin{document}
\baselineskip=16pt
\begin{titlepage}

\begin{center}
\vspace*{12mm}

{\Large\bf%
Holographic Multi-Band Superconductor
}\vspace*{10mm}

Ching-Yu Huang$^{1}$\footnote{e-mail:
{\tt 896410093@ntnu.edu.tw}},
Feng-Li Lin$^{1}$\footnote{e-mail:
{\tt linfengli@phy.ntnu.edu.tw}}
and
Debaprasad Maity$^{2,3}$\footnote{e-mail:
{\tt  debu@phys.ntu.edu.tw}}
\vspace*{5mm}

${}^1$
{\it Department of Physics, National Taiwan Normal University,
Taipei 116, Taiwan,
\\[3mm]
${}^2$
Department of Physics and Center for
Theoretical Sciences, National Taiwan University, Taipei 106, Taiwan\\
${}^3$ Leung Center for Cosmology and Particle Astrophysics\\
National Taiwan University, Taipei 106, Taiwan
\\[3mm]
}

\end{center}
\vspace*{10mm}

\begin{abstract}\noindent%

\vspace{3ex}


\end{abstract}
   We propose a gravity dual for the holographic superconductor with
multi-band carriers. Moreover, the currents of these carriers are
unified under a global flavored SO(3) symmetry, which is dual to the
bulk SO(3) gauge symmetry. We study the phase diagram of our
model, and find it qualitatively agrees with the one for the realistic
2-band superconductor, such as $MgB_2$. We also identify the bulk field dual to the electromagnetic $U(1)_{EM}$ current, which should be invariant under the global flavored SO(3) rotation. We then evaluate the corresponding holographic conductivity and find the expected mean field like behaviors.

\end{titlepage}

\newpage

\section{Introduction}

  Symmetry principle is believed to be one of the
important guiding principles in constructing the new physics models.
As is well known in the context of standard model of particle physics,
some non-trivial dynamics could be unified in the form of non-Abelian
symmetry, such as gauge symmetry for electroweak and strong interactions
or the approximate global flavor symmetry of quarks.
In contrast, the symmetry-unified dynamics in the condensed matter
physics is less explored and appreciated.
Despite that, there was an exceptional SO(5) model proposed by Zhang
\cite{Zhang} as the unified model of d-wave superconductivity (d-SC) and
 anti-ferromagnetism (AF).
However, in order to explain the phase diagram of high temperature (high $T_c$) superconductivity  one needs to add the explicit symmetry breaking
terms in this model.

   In the context of AdS/CFT correspondence, the global symmetry
of the boundary CFT is dual to a gauge symmetry of the bulk
gravity. For example, in the study of meson dynamics of the
holographic QCD, the flavor symmetry of QCD is dual to the gauge
symmetry on the probe mesonic branes \cite{SS model}. Similarly,
in the recent proposal of holographic superconductor
\cite{Hartnoll:2008vx,Hartnoll:2008kx}, the global U(1) symmetry for the
charge conservation is again dual to a U(1) gauge symmetry in the
bulk gravity. These gravity models indeed capture the essential
feature of the dual CFTs.

 The gauge symmetry usually constrains on
the dynamics more than the global one can do. In the context of AdS/CFT,
this could imply that the constraint on the dynamics due to a speculated bulk gauge symmetry
might uncover some emergent IR phenomenon of the dual CFT.
One can then imagine that some originally disjoint flavor symmetries
could be unified into a non-Abelian one at low energy due to the
direct or indirect couplings among the various flavor currents.
A typical example is the aforementioned SO(5) superconductor, in which
the symmetry could be thought as an enlarged emerging symmetry
at low energy by merging AF's SO(3) and d-SC's U(1). Motivated by
this model and the emerging symmetry principle, it is then tempting
to formulate a holographic model of high $T_c$ superconductivity based
on some underlying non-Abelian gauge symmetry.

   Naively, we can consider the gauge fields and the fundamental scalar
under SO(5) gauge symmetry in the AdS-Schwarzschild background
as the holographic dual to Zhang's unified theory of high temperature
superconductors.  We may then ask if it is possible for the black hole to grow
the non-Abelian hairs by tuning the
asymptotic values of the gauge fields, i.e., the corresponding
chemical potentials for the dual flavor charge carriers.
If so, one may wonder if it is possible to reproduce the
peculiar phase diagram of the high $T_c$ superconductors.
However, a quick thought will turn down the proposal. This is
 because the phase diagram of the high Tc superconductor shows
the competition between the  anti-ferromagnetic order and the
d-wave superconducting order, which is in conflict with the picture
of coherent orders dictated by the underlying gauge symmetry.
Here, by coherent orders we mean that different order parameters
will influence each other to condense at the same temperature.
This is in clear contrast with the competing
order phenomena in high Tc superconductivity.
Indeed, in \cite{Zhang} one needs to add the explicit
SO(5) breaking terms in order to achieve the phase diagram with competing orders. Translated into the  gravity dual picture, one needs to break the SO(5) gauge symmetry
explicitly, which will usually lead to
 inconsistence as for massive gauge theories not via Higgs mechanism.

  We will not consider the complicated broken SO(5) case. As a first step,
we believe that the superconductivity with coherent orders is also an 
interesting physical phenomena  to look into. We can introduce a 
non-Abelian  gauge symmetry in the bulk to describe coherent 
orders of the boundary field theory. 
The coherent orders are arising from the condensation of the different kind of 
charge carriers.  In the holographic QCD, these non-Abelian gauge fields are 
well known to be holographically dual to the quark or meson flavors.
Similarly, we will interpret  different component of the non-Abelian 
gauge field to be the holographic dual to some current associated with 
the different band carriers. From the usual AdS/CFT dictionary, asymptotic
boundary values of the gauge fields will be dual to the
chemical potentials of the corresponding band carriers.
Unified non-Abelian symmetry can be
understood as an emergent global symmetry
due to the nontrivial interactions among the different band carriers.
This is the origin of the coherent orders. 

 It is interesting to mention that there indeed exist
the real multi-band superconductors such as Magnesium diboride ($MgB_2$)
\cite{MgB2-nature,MgB2} and the recently discovered iron pnictides \cite{iron}.
The $MgB_2$ does show the coherent orders in its
superconducting ground state. More specifically, it has
two energy gaps for two different band carriers with the same critical temperature
$T_c$. However, there is no clear physical understanding why the coherent orders occur.  
Our study provides a clue to explain it by the underlying non-Abelian symmetry. 

    In this paper, we will study the most simplest non-Abelian
symmetric holographic multi-band superconductor, namely the
model based on a bulk SO(3) gauge symmetry. The appearance of the
multiple energy gaps with the same $T_c$ is an interesting outcome
of this paper. Our results also show that a sub-sector of this model
reproduce the coherent orders of the 2-band superconductor.
This may imply that the underlying dynamics of $MgB_2$ superconductor has
a hidden SO(3) symmetry at low energy. Beside these, the phase diagram
for the full 3-band case also shows interesting feature.
At this point we would like to remind the readers
that our model is different from holographic p-wave
superconductors considered in \cite{Gubser:2008wv}, where only
the non-Abelian gauge fields are introduced, not the fundamental scalars.

   We would like to emphasize that the motivation for the holographic study of
 the condensed matter systems is to uncover the universal behaviors of the systems, especially
 the ones dictated by the underlying symmetries. For examples, the U(1) symmetry is important for
 the holographic s-wave superconductor \cite{Hartnoll:2008vx} to have the mean-field like behavior, the SU(2) symmetry  is crucial for the holographic p-wave superconductor \cite{Gubser:2008wv}, and the near horizon 1+1 conformal symmetry can explain the peculiar behavior of the holographic non-Fermi metals \cite{Faulkner:2011tm}. Again our holographic SO(3) model can explain the appearance of the coherent orders. This is also beyond the scope of the holographic U(1) model. Therefore, our result implies that the appearance of the coherent orders are universal in the system with non-Abelian symmetry among different band carriers.

   The coherent orders in the real multi-band materials such as $MgB_2$ are believed to be due to the phonon coupling between the different band carriers, and can be obtained from the first principle calculation based on some microscopic models \cite{multibandth}. However, in this kind of approach it usually needs fine-tuning of the model parameters, and lacks the physical understanding of the underlying dynamics. In this regard, our SO(3) holographic model provides the physical insight for the underlying dynamics of the coherent orders, namely, it is dictated by the emergent SO(3) symmetry among  the different band carriers. Though our result holds for the strong dynamics based on its gravity dual, it should also hold for the weak coupling regime as long as the same SO(3) symmetry emerges there. The strength of the coupling will only affect the mean-field behavior quantitatively, but not qualitatively. Indeed, we will see this is the case as in the holographic U(1) superconductor. Our finding may trigger in searching for the new strongly correlated materials which  are unconventional superconductor with multiple order parameters and have the same critical temperature.

    Another interesting challenge for our proposal is how to 
identify the physical electromagnetic $U(1)_{EM}$ in order to 
calculate the holographic conductivity. 
Naively, the $U(1)_{EM}$ could be the Cartan sub-algebra of SO(3). However, 
there is no unique choice since the arbitrary proper linear combination of the 
SO(3) generators will play the equivalent role.  We need to 
find some additional criterion for such an identification.  In analogy to QCD, 
the flavored SO(3) current is usually different from the 
electric current, and the latter should be invariant 
under the rotation associated with the flavor symmetry. In this way, 
we will identify the bulk field which is dual to the $U(1)_{EM}$, and 
then evaluate the corresponding holographic conductivity.

     The paper is organized as follows. In the next section we will
 pull out the equations of motion for the gauge fields and fundamental
 scalars, and give proper holographic interpretation. In section 3 we
 numerically solve the equations of motion for the background fields
in the probe limit, and display the phase diagrams for the holographic
 multi-band superconductor. In section 4 we identify a gauge-invariant Cartan sub-algebra as the physical U(1) coupled to photon, and evaluate the corresponding holographic conductivity. Finally we briefly conclude our paper in section 5. In Appendix, we give the numerical results for the SO(3) conductivity matrix.

\section{SO(3) in AdS-Schwarzschild background}

  As we have already mentioned in the Introduction, we
will consider SO(3) gauge fields and fundamental scalars
in the AdS-Schwarzschild black hole background.
The scalars are the holographic duals to the superconducting order
parameters of the boundary theory. The dual boundary global
symmetry can be thought of as enlarged unified symmetry of
multiple U(1) order parameters of
some superconductors with multi-band carriers, e.g., the 2- or 3-band
superconductors. Microscopically, the unification of the symmetry could
arise from the indirect interaction among the different band carriers
via the phonon coupling. As we will see, our model reproduces the
coherent feature of the order parameters for the 2-band superconductors,
and this may justify the hidden SO(3) symmetry of the underlying unified
dynamics for the 2-band carriers.

  The action for our holographic multi-band superconductor model is
\begin{equation}\label{Maction}
S=\int
d^4x\sqrt{-g}(R+\frac6{L^2}-\frac18TrF^2_{\mu\nu}-|D_\mu \phi|^2-m^2\phi^2)\;,
\end{equation}
where the scalar $\phi$ is in the fundamental representation of SO(3), i.e.,
$\phi=(n_3,n_2,n_1)^T$, and the covariant derivative $D_{\mu}\phi:=
\partial_{\mu}\phi-i q A_{\mu} \phi$. $q$ is the Yang-Mills
gauge coupling parameter. The gauge connection $A_{\mu}$
is in the adjoint representation, i.e.,
\begin{equation}\label{generators}
A_{\mu}=i\left( \begin{array}{ccc}
        0 & -A^1_{\mu} &-A^2_{\mu} \\
        A^1_{\mu} & 0 & -A^3_{\mu} \\
        A^2_{\mu} & A^3_{\mu} & 0  \\
         \end{array} \right)
         \equiv \sum_{i=1}^{3}A^i_{\mu}\tau^i\;,
\end{equation}
where the $\tau^i$'s are hermitian SO(3) generators obeying the
 SO(3) Lie algebra, i.e.,
\begin{equation}
[\tau^i~\tau^j]=if^{ijk}\tau^k\;, \qquad tr(\tau^i\tau^j)=2\delta^{ij}\;.
\end{equation}
Here $f^{ijk}$'s are the structure constants of the SO(3) Lie-algebra, i.e.,
 $f^{123}=f^{231}=f^{312}=1$, etc. The explicit representations of
$\tau_i$'s can be read from \eq{generators}. The field strength
 is then given by
\begin{equation}
F^i_{\mu\nu}\equiv\partial_{\mu}A^i_{\nu}-
\partial_{\nu}A^i_{\mu}+ q f^{jki}A^j_{\mu}A^k_{\nu}
\end{equation}
or in more compact form $F_{\mu\nu}\equiv F^i_{\mu\nu}\tau^i=
\partial_{\mu}A_{\nu}-\partial_{\nu}A_{\mu}-i q [A_{\mu}~A_{\nu}]$.

The gauge field $A^i_{\mu}$ is the holographic dual of the current
 $J_{\mu}^i$ consisting of the carriers in the $i$-th band, and the
 scalar field $n^i$ is dual to the mean field order operator
 $O^{(i)}$ in the $i$-th band. The SO(3) symmetry is the
aforementioned unification of three U(1) bands of the
carriers due to some microscopic dynamics such as phonon
coupling.  From the action \eq{Maction} we can derive the
equations of the motion. The equation of motion for $\phi$ is
\begin{equation}
\frac1{\sqrt{-g}}\partial_{\mu}(\sqrt{-g}D^{\mu}\phi)-
iqA^{\mu}D_{\mu}\phi-m^2 \phi=0
\end{equation}
The equation for $A^i_{\mu}$ is
\begin{equation}\label{YM}
{1\over \sqrt{-g}} \partial_{\mu}(\sqrt{-g} F^{i \mu\nu})+
q f^{ijk}A^{j}_{\mu}F^{k \mu\nu}=iq[ \phi^{T} \tau^i
 D^{\nu}\phi - (D^{\nu}\phi)^{T} \tau^i \phi]\;.
\end{equation}

   To mimic the dual superconductor, we should put the probe
 gauge fields and scalar on a bulk black hole background with
 the standard AdS-Schwarzschild metric
\begin{equation}
ds^2=-f(r)dt^2+\frac{dr^2}{f(r)}+r^2(dx^2+dy^2)
\end{equation}
where $f(r)=\frac{r^2}{L^2}-\frac{M}r$. As usual, the temperature
 of the black hole is $T={3M^{1/3} \over 4\pi L^{4/3}}$,
 which is also the temperature of the dual boundary theory.

 We now consider the probe background gauge fields as following
\be\label{aansatz}
A_{\mu}dx^{\mu}:=[ \Pi_1(r)\tau^1+\Pi_2(r) \tau^{2} +\Pi_3(r) \tau^3] dt\;,
\ee
and the background scalar field configuration
\be\label{scalar1}
\phi:=(n_3(r),n_2(r),n_1(r))^T\;.
\ee
 Note that the ordering is the reverse of the conventional
one to make its compatible with the labeling for the gauge field.

  The equations of motion for $n_i(r)$ are
\be \label{eomn} n''_i+({f'\over f} +{2\over r})n'_i+
{q^2\over f^2}(\Pi n)_j {\partial (\Pi n)_j
\over \partial n_i}-{m^2 \over f} n_i=0\;,
\ee
for $i,j=1,2,3$. In the above we have defined the bracket
vectors as follows,
\be
(AB)_1=-(A_1B_2+A_2B_1)\;, \qquad (AB)_2=A_1B_3-A_3B_1\;,
 \qquad (AB)_3=A_2B_3+A_3B_2\;. \label{bracketv}
\ee

 The equations of motion for $\Pi_i(r)$ are
\be 
 \Pi_i''+{2\over r}\Pi_i'-{2q^2\over f}(\Pi n)_j {\partial
 (\Pi n)_j \over \partial \Pi_i}=0\;, \label{eompi}
\ee
along with the first order gauge constraints  due to our ansatz \eq{aansatz}
\bea\label{cs1}
&& {\cal R}_1= (\Pi_2\Pi'_3-\Pi_2'\Pi_3)- 2  f (n_3n'_2-n'_3n_2)=0 \nno;\\
&& {\cal R}_2= (\Pi_1\Pi'_2-\Pi'_1\Pi_2)-2 f (n_2n'_1-n'_2n_1)=0\;,\\
&& {\cal R}_2=(\Pi_3\Pi_1'-\Pi'_3 \Pi_1)-2  f (n_3n'_1-n'_3n_1)=0\nno;. 
\eea
In order to solve the equations of motion, we require the above
 gauge constraints to be consistent with the equations of motion.
 This will yield a consistent set of boundary conditions for the gauge fields
 and scalar fields at horizon. To see this, we consider the
proper combination of the field equations (\ref{eomn}) and (\ref{eompi}). 
One can easily see that out of the combinations of these six equations, there are  three independent equations as follows
\bea 
{\cal R}'_i + \frac {{\cal R}_i} r =0  \implies {\cal R}_i = \frac {C_i} {r^2}
\eea
where $i=1,2,3$, and $C_i$'s are the integration constants. So, in order to be
consistent with the constraint equations (\ref{cs1}), the above three integration constants
should be zero. This will impose the boundary conditions for the gauge and scalar fields at the black hole 
horizon. One such a consistent set of boundary conditions would be to have vanishing  gauge fields
and regularity of the scalar fields at the horizon. Interestingly, this is precisely the set of boundary condition which we
usually consider in the holographic set up. With these above mentioned choice of conditions, we can solve the equation of motions without worrying about the constraint equations.

 At this point, it is important to note that we can
reduce this 3-band model to a 2-band
one by setting one of the following pairs to zero,
$(n_1,\Pi_1)$, $(n_2,\Pi_2)$ or $(n_3,\Pi_3)$.  We can further reduce to the familiar U(1) model by setting two of the above pairs to zero. These relations imply that the multi-band models are deeply related to the U(1) case. We will see this is indeed the case by the similarity of the phase diagrams.

  On the other hand, we cannot reduce a 3-band configuration to a 2- or 1-band configuration by gauge transformation. The reason is that the vacuum is Higgsed and the gauge symmetry is broken so that the configurations connected by the  gauge transformation will have different energies, and cannot be physically equivalent. This can be checked explicitly by showing that one cannot reduce the number of non-zero gauge fields and the scalars for a given configuration at the same time by gauge transformation. This holds even we just perform the global transformation at asymptotic infinity.

\section{Phase Diagrams}
In this section, we will solve the equations of motion \eq{eomn}-\eq{eompi} by the numerical shooting method, and find out the phase diagrams. Note that there are 6 functions to be solved so that we write a Fortran program of shooting method to perform such a task.

  As usual, we need to impose the boundary conditions to
solve the equations of motion. Moreover, the chosen boundary conditions should be also consistent with the gauge constraints \eq{cs1}. After manipulating the combinations of the gauge constraints,
we find that the consistent boundary conditions are pretty much the same as the ones for the holographic U(1) superconductor with vanishing gauge fields
and regularity of the scalar fields at the black hole horizon, namely, at the horizon $r=r_0$,
$\Pi_i=0$ so that $\Pi_i dt$ has finite norm, and
the equations of motion for $n_i$ implies $n_i=3r_0 n_i'/m^2L^2$.

  Hereafter, we will choose $L=1$ and $m^2=-2$ so
that $n_i$ is dual to a CFT operator ${\cal O}^{(i)}$
with conformal dimension $1$ or $2$. This yields the
 following asymptotic behaviors at $r=\infty$,
\bea
n_i&=&{n_i^{(1)}\over r}+{n_i^{(2)} \over r^2}+ \cdots\;,\\
\Pi_i &=&\mu_i -{\rho_i \over r}+\cdots\;.
\eea
 From the above we can read off the properties of the
dual CFT, i.e., the condensate of the operator ${\cal O}^{(i)}$ is given by
 \be
 \langle {\cal O}^{(i)}_a \rangle =\sqrt{2}n_i^{(a)}, \qquad  a=1,2
 \ee
with $\epsilon_{ab} n^{(b)}_i=0$. For simplicity, we only
 consider the case of $a=2$ in this paper. The value of
 $\langle {\cal O}^{(i)}_2 \rangle$ at zero temperature
is the energy gap for the $i$-th band carrier to form the
 BCS-like Cooper pairs, and the values of $\mu_i$ and
 $\rho_i$ are the chemical potential and the carrier
 density of the $i$-th band carriers, respectively.

\begin{figure}[ht]
\center{\epsfig{figure=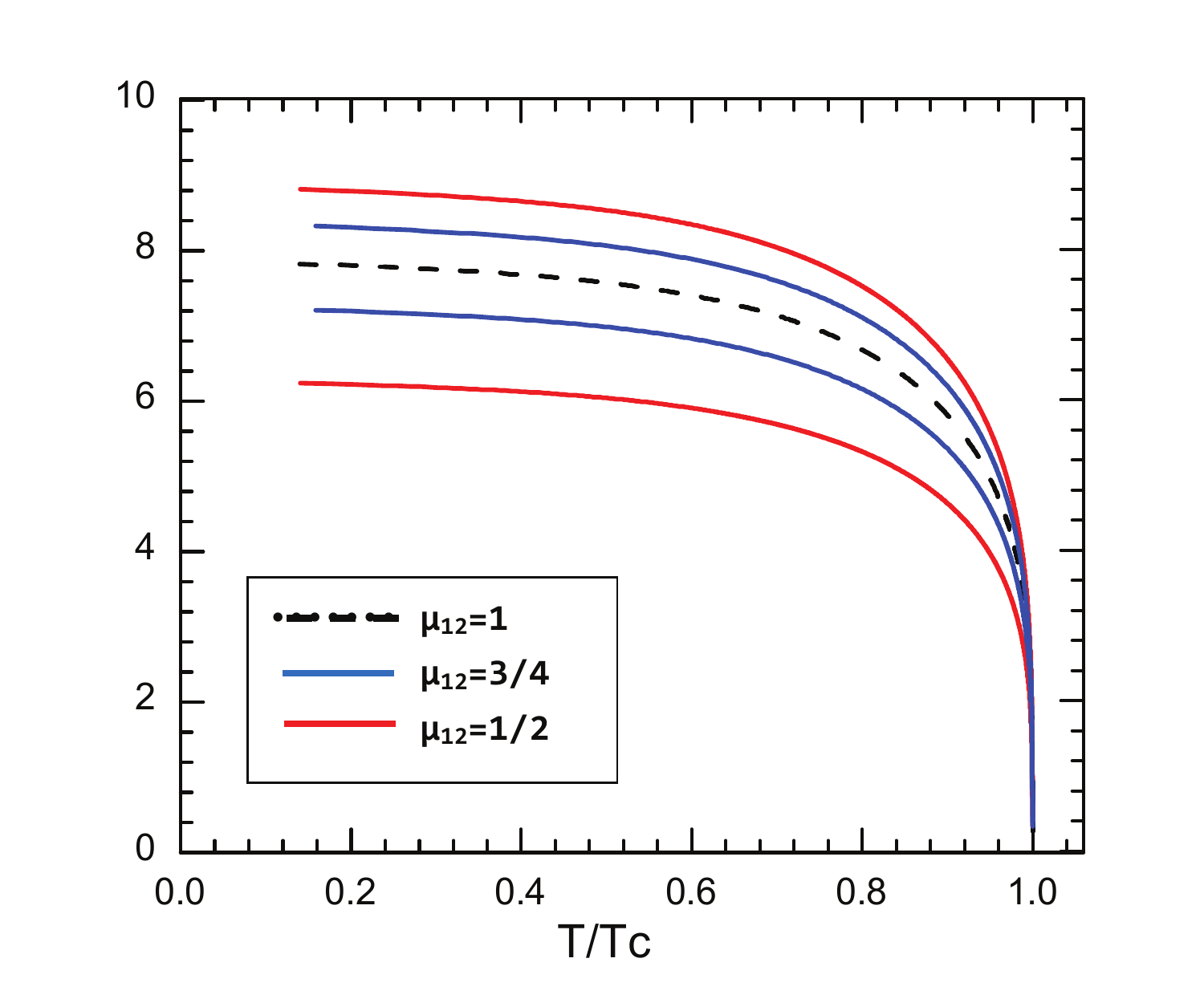,angle=0,width=8cm}}
\caption{${\sqrt{\langle {\cal O}^{(i)}_2 \rangle} \over T_c}$ v.s.
${T\over T_c}$, $i=1,2$ --- the phase digram for the holographic
 2-band superconductor, which is only a function of
$\mu_{12}:={\mu_1\over \mu_2}$.
 We show the cases with $\mu_{12}=1,3/4,1/2$. }
\label{2-band}
\end{figure}

    Since the  2-band superconductor is well studied by the experiments,
we will first focus on the the 2-band case of our model by setting
$n_3=\Pi_3=0$, i.e., we will consider only the pairs $(n_1,\Pi_1)$
and $(n_2,\Pi_2)$. The phase diagram from our numerical result is
shown in Fig. \ref{2-band}. We see that the phase diagram in terms
of the dimensionless quantities is universal, and is only function
of $\mu_1/\mu_2$. More importantly, the phase diagram shows coherent
 orders, and each order parameter obeys the BCS-like universal
scaling behavior, i.e., the carriers of 2 bands condense at the
 same $T_c$ with the universal critical behavior as the real
MgB2 2-band superconductor does \cite{MgB2}.  By the numerical
fitting, we find

\be
\langle {\cal O}^{(i)}_2 \rangle \simeq 163 T_c^2
{\mu_i \over \sqrt{\mu_1^2+\mu_2^2}} (1-{T\over T_c})^{1/2}\;,
 \qquad i=1,2  \qquad \mbox{for}\;\; T \simeq T_c\;.  \label{2-band O}
\ee
This is in contrast to the case for U(1) holographic
superconductor, $\langle {\cal O}_2 \rangle \simeq
144 T_c^2  (1-{T\over T_c})^{1/2}$.  However, in both
 cases we all have the mean-field critical exponent $\beta=1/2$.

    Moreover, the scaling relation between $T_c$ and the
carriers' densities is as follows from the numerical fitting
\be
T_c\simeq 0.118 \sqrt{\rho_1^2+\rho_2^2}.
\ee
This is in analogy to the one for the U(1) case, i.e.,
 $T_c \simeq 0.118 \rho^{1/2}$.

  Up to now, our results of the phase diagram agree well with
 the BCS-like behavior for the 2-band superconductor, it
 could be the strongly coupled version of the ordinary 2-band superconductor.

\begin{figure}[ht]
\center{\epsfig{figure=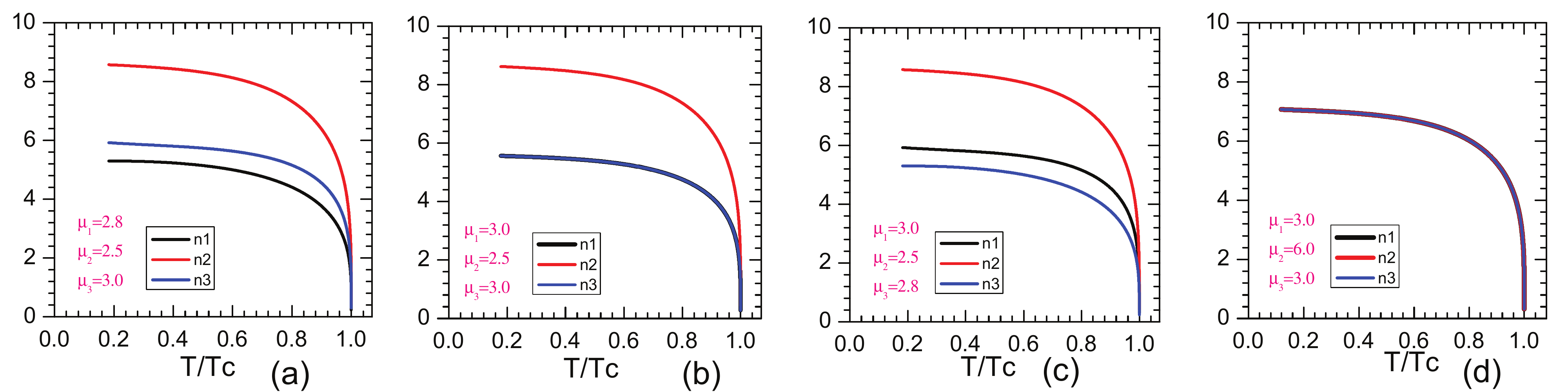,angle=0,width=18cm}}
\caption{${\sqrt{\langle {\cal O}^{(i)}_2 \rangle} \over T_c}$ v.s.
 ${T\over T_c}$, $i=1,2,3$ --- the phase digram for the holographic
3-band superconductor. We show the band-gap competition by tuning
the chemical potentials. }  \label{3-band}
\end{figure}

   Now we turn on all 3 bands at the same time, numerically
we find that the results are more sensitive to the intrinsic
 numerical errors as expected. The phase diagrams show the
similar mean-field feature as the 2-band case. Some of the
typical phase diagrams are shown in Fig. \ref{3-band}. It
 is interesting to see that by tuning the chemical potentials
$\mu_i$, one can collapse the 3 bands into 2-band or 1-band
cases. Moreover, there is a inversion of the vevs or the
 energy gaps as shown in Fig. \ref{3-band}(a) and (c).

   As noted, the temperature dependence of the $\langle {\cal O}^{(i)}_2
\rangle$'s near the critical point still conforms to  the mean-field behavior,
however, the chemical potential dependence is far more complicated
than \eq{2-band O} for the 2-band case.  We cannot find the
complete dependence on the chemical potentials, but just the
 proportionality relation as follows.
\be
\langle {\cal O}^{(i)}_2 \rangle = C_i T_c^2 (1-{T\over T_c})^{1/2}\;,
\qquad i=1,2  \qquad \mbox{for}\;\; T \simeq T_c\;,
\ee
with
\be
C_1:C_2:C_3={2\mu_2 |\mu_3-\mu_2| \over |\mu_1-\mu_2|+|\mu_3-\mu_2|}:
 \mu_1+\mu_3: {2\mu_2 |\mu_1-\mu_2| \over |\mu_1-\mu_2|+|\mu_3-\mu_2|}\;.
\ee
On the other hand, the scaling relation between the critical
temperature and the carriers' densities has the similar form as
the 2-band and the $U(1)$ cases, namely,
\be
T_c\simeq 0.118 \sqrt{\rho_1^2+\rho_2^2+\rho_3^2}.\ee

So far we have discussed about the superconducting phase
diagram of a holographic multi-band superconductor. Behavior of the
condensation for each band is universal in terms of temperature.
As we decrease the temperature, condensation happens at the 
same temperature for all the band carries. As we have mentioned
in the introduction this phenomena is quite similar to the 
real two-band superconductor $MgB_2$. In the subsequent section we
will calculate the superconducting transport properties like optical 
conductivity under small electromagnetic perturbation.  

\section{Holographic Conductivity}
In this section, we would like to derive the field equations for the gauge
 field perturbation on the above background. We then solve these
 equations to  obtain the holographic real time Green functions
of boundary currents, from which we can extract the conductivities.

  Let us turn on the gauge field perturbation of the $x$-component
 as follows,
 \be\label{pertA}
\delta A_{\mu}dx^{\mu}:=e^{-i(\omega t-k_2 y)} [ a_1(r)\tau^1+a_2(r)
\tau^{2} +a_3(r) \tau^3] dx\;,
\ee

   Plugging the perturbed fields into the field equations \eq{YM},
 and expand it up to the linear order, from which we can derive the
 field equations for the perturbed fields. The results are the
following,
\be\label{YMp}
 a_i''+{f'\over f}a_i'+{1\over f^2}\left(\omega^2 a_i+2iq\omega
f^{ijk}\Pi_j a_k +q^2 (\vec{\Pi}\cdot\vec{\Pi}a_i-\Pi_i
 \vec{\Pi}\cdot \vec{a})\right)
 -{1\over f} \left(2q^2 (an)_j {\partial (a n)_j \over
\partial a_i} +{k_2^2 \over r^2} a_i \right)=0\;,
\ee
where we define $\vec{\Pi}:=(\Pi_1,\Pi_2,\Pi_3)$, $\vec{a}:=(a_1,a_2,a_3)$,
  and the inner product such as $\vec{\Pi}\cdot \vec{a}=\Pi_1 a_1+\Pi_2 a_2+\Pi_3 a_3$.
In the above expressions, we have again used the bracket notation
defined in \eq{bracketv}. We also have noticed that there
is no constraint equations among $a_i$'s perturbations as long as 
we consider a specific form (\ref{pertA}) of the electromagnetic perturbation.
Therefore, from the holographic point of view the external electric field at the boundary does not produce any  scalar perturbation in our background.

   Similar to the abelian case, we then solve the
 above equations by imposing the incoming wave boundary condition
in the near horizon region, namely
$a_i=f^{-i\omega L^2/3r_0} [1+ a_{i,1}(r-r_0)+\cdots]$.
Then from the asymptotic behavior of all the fields at the asymptotic
boundary we found 
\be\label{asymp}
 a_j(r,\vec{k},\omega)=a_j^{(0)}(\vec{k},\omega)+{a_j^{(1)}\over r}
(\vec{k},\omega)+\cdots ,
 \ee
 It then seems that one can evaluate the holographic conductivity by the standard holographic prescription for the Ohmic law  \cite{Son:2002sd,Iqbal:2009fd,Iqbal:2008by,Hartnoll:2007ip}.  
However, one should be careful about the fact that all the perturbed fields $a_i$'s are linearly coupled to each other in the bulk.  This coupling among the fields implies that source of one particular perturbed field will also  source the current corresponding to the other components.  If we identify $a_i$ as dual to the U(1) current for the $i$-th band, this will then yield
\be \label{condten}
\langle J_{i}\rangle = \sigma_{ij} E_j,
\ee
with $3\times3$ matrix $\sigma_{ij}$ as a general conductivity matrix 
in the global SO(3) space. Noe that the $i,j$ are the indices for the SO(3) internal space, not the space-time ones. Since the conductivity matrix cannot be measure directly, we will not discuss it further here but in the Appendix. 

In our model we have three different kinds of carriers 
corresponding to three components of a fundamental scalar field in 
the bulk. Each carrier has the corresponding boundary current dual to $a_i$ . 
This dual current transforms like a vector under the global 
internal symmetry. This is, however, not the current that we are 
interested in. Now how do we define the conductivity under U(1) 
electromagnetic field? According to the standard holographic prescription, 
the above currents are actually some kind of weekly gauged flavour 
currents similar to the standard s-wave superconductor. 
As we have argued, the system we are considering is a holographic
multi-band superconductor. So, all the band carriers should 
couple to a single physical photon field through which we will 
calculate the optical conductivity of the system. 

Naively, we can identify one of the three generators of SO(3) 
as the physical $U(1)_{EM}$, namely the Cartan subalgebra, 
and the other two as the ladder operators. More specifically,
  \bea
[\tau^3,\tau^{\pm}] = \pm \tau^{\pm}, ~~~ [\tau^+, \tau^-]= \tau^3.
\eea
However, the above identification is not unique, and any linear 
combination of the three SO(3) generators with proper 
normalization will play the same role.  
Even though there is
an arbitrariness in defining the U(1) subgroup, 
any physical quantity should be independent
of that particular choice. But this does not happen in our case. 
Specifically, considering above particular definition, we can 
find out the solution for $a_3$ which coupled to a boundary
current with the source of physical electric field, 
but other two components $a_1$ and
$a_2$ are sourceless or normalizable . 
Then one can calculate $\sigma_{33}$ and identify it as a physical
conductivity using the standard holographic prescription. 
This is how we indeed calculated the
conductivity matrix as shown in the Fig.\ref{condmatx} of the Appendix. We can apply the same 
procedure for the other choice of U(1), 
but those choices lead to different values of the conductivity as one
can easily see from the plot for the conductivity matrix. 
This discrepancy is intuitively obvious because the
background chemical potential for each band carrier is different. 
So, when we consider different
choice of U(1) subgroup of the bulk gauge group, 
we are actually not considering same photon field at the boundary.

Therefore, the Cartan subgroup cannot be the physical $U(1)_{EM}$ 
for the electromagnetism. Instead, the physical U(1) electric 
current should be independent of the choice of the Cartan subalgebra,
invariant under the global SO(3) rotation and also does not 
depend on the intrinsic properties of the individual band carrier. 
We therefore can interpret 
that particular combination as a dual
to the physical $U(1)_{EM}$ current at the boundary. 
Accordingly, we can evaluate the holographic
conductivity which is invariant under the SO(3) 
transformation and also does not depend on
individual chemical potential. 

  It seems strange that the physical $U(1)_{EM}$ current is defined only for the perturbation but not for the background. However, in the holographic approach this is natural because the background gauge field is dual to the chemical potential and carriers' density, which define the properties of the vacuum not the dynamics. In fact, the different band carriers should have the corresponding chemical potentials and carriers' densities as specified by the background profiles.  In contrast, the perturbation of the gauge field is dual to the electric source and the dynamical current in the context of linear response theory. In this case, all the carriers are coupled to the same $U(1)_{EM}$ source, and an unambiguous identification of physical $U(1)_{EM}$ is necessary.

   Mapping the above consideration back to the bulk point of view, 
the SO(3) rotation is translated into the gauge transformation 
for the total gauge field, namely,  
\be
\delta A^i_{\mu}=\partial_{\mu} \alpha^i + q f^{ijk}A^j_{\mu} \alpha^k 
\ee
where the total gauge field $A^i_{\mu}$ includes both the  the background ansatz \eq{aansatz} and the fluctuation one \eq{pertA}, and $\alpha^i$'s are the gauge functions. Since we are only interested in the SO(3) rotation relating different choices of Cartan U(1), this implies that the gauge function functions should be independent of the boundary coordinates. This particular class of gauge transformation retain the transformed total gauge field in the same class\footnote{Namely, the background is a time-component of the gauge field as only a function of $r$, and the perturbation is the $x$-component of the gauge field.} of the ansatz \eq{aansatz} and \eq{pertA} if we also require the gauge functions are also independent of $r$. Then, the gauge functions become constant gauge parameters and the gauge transformations reduce to the global rotation, and the only nontrivial parts of the transformations are
\be\label{so3}
\delta \Pi_i=f^{ijk}\Pi_j \alpha^k\;, \qquad \delta a_i=f^{ijk}a_j \alpha^k.
\ee
      
\begin{figure}
\center{\epsfig{figure=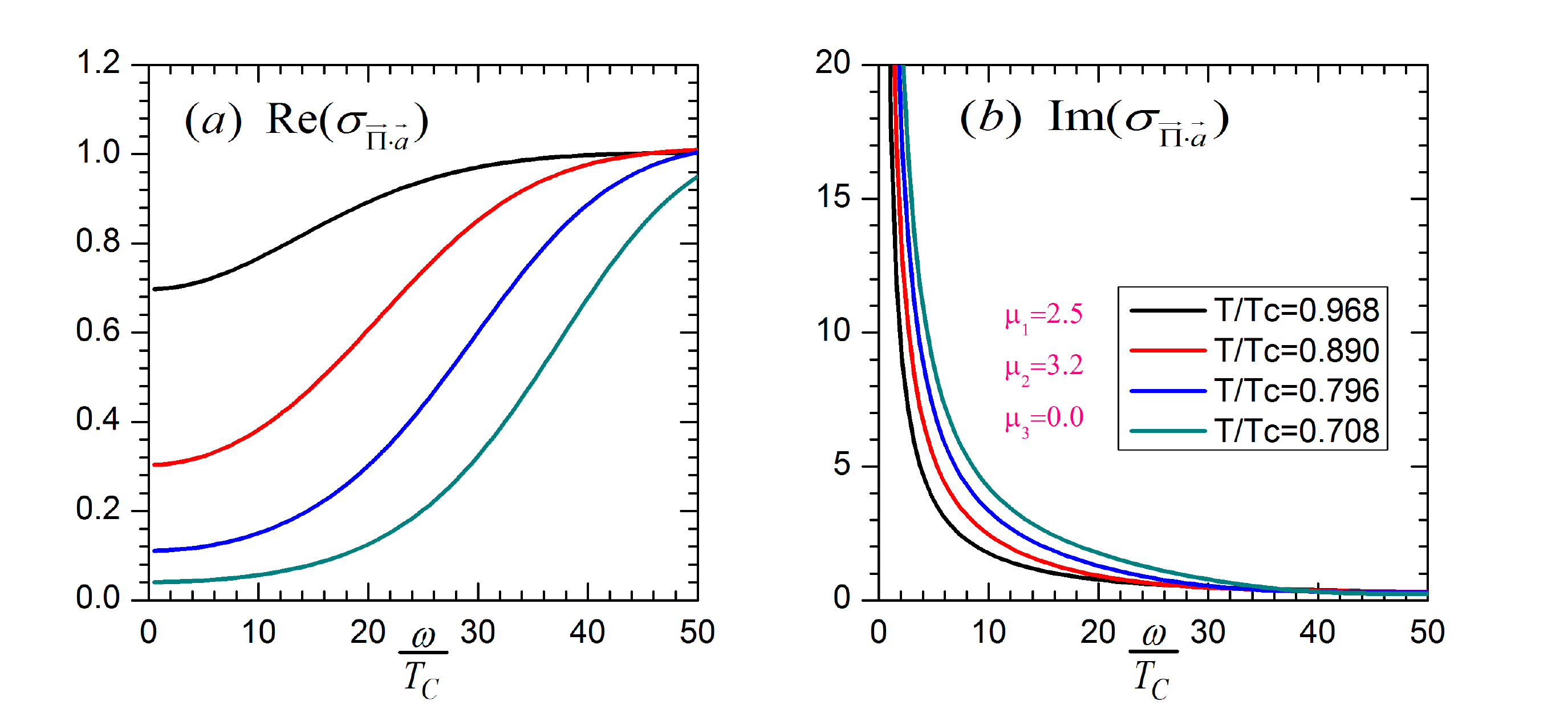,angle=0,width=15cm}}
\caption{The AC conductivity $\sigma_{\vec{\Pi}\cdot \vec{a}}$
of 2-band superconductor for $\mu_1=2.5,\mu_2=3.0,\mu_3=0.0$.
(Left): the real part of conductivity,
and (Right): the imaginary part.}  \label{n1n2}
\end{figure}      
      
   The task to find the dual field of physical $U(1)_{EM}$ is then equivalent to find the appropriate linear combination of the gauge fields $a_i$'s so that it is invariant under the transformation \eq{so3}. it is then straightforward to see the field $\vec{\Pi} \cdot \vec{a}$ satisfies this constraint, and we will identify it as the holographic dual to the physical $U(1)_{EM}$ current.  The peculiar dependence of this dual current on the background profile $\vec{\Pi}$ may reflect the physical relevance that the current here is actually the vev of the current operator, which will then depend on the properties of the vacuum encoded in $\vec{\Pi}$.

 \begin{figure}[ht]
\center{\epsfig{figure=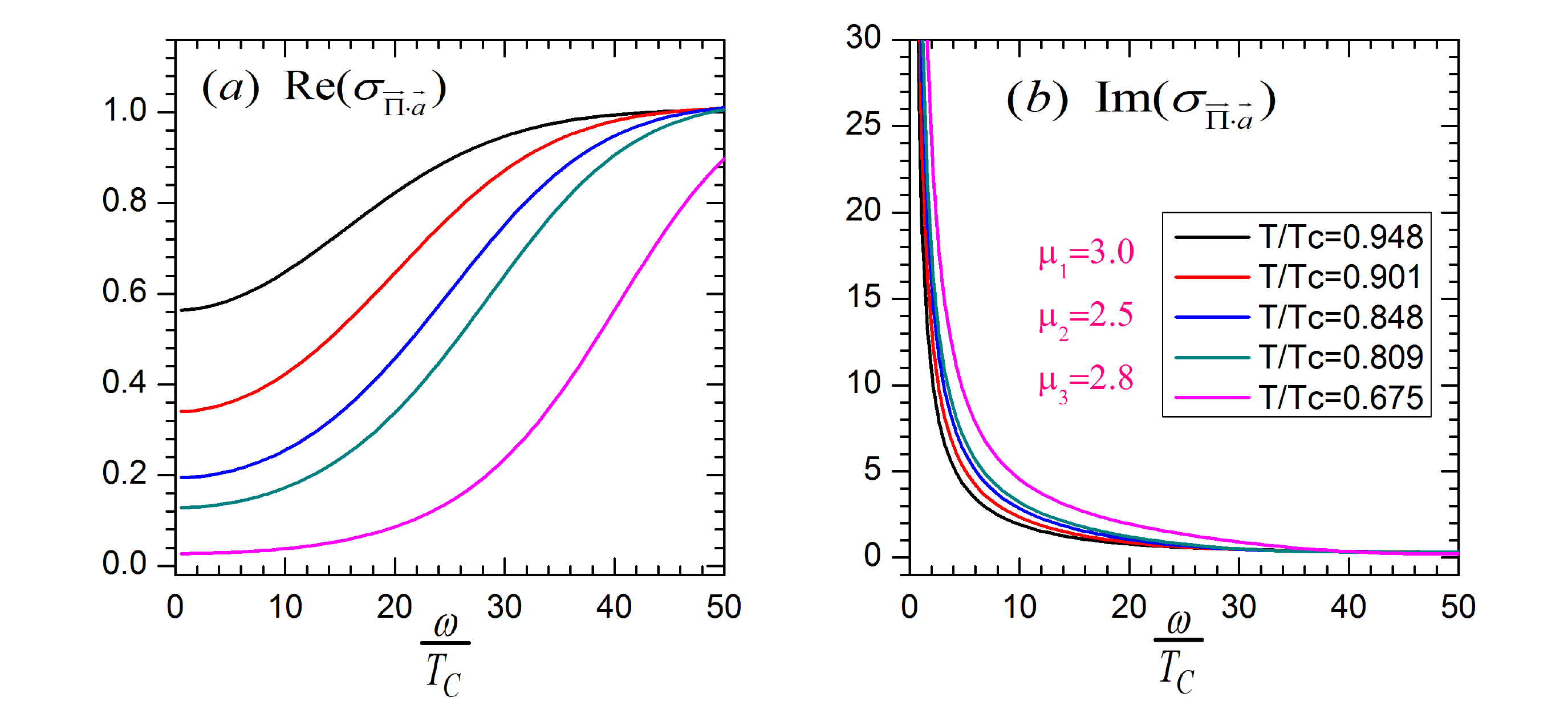,angle=0,width=15cm}}
\caption{The AC conductivity $\sigma_{\vec{\Pi}\cdot \vec{a}}$ of 3-band superconductor for $\mu_1=3.0,\mu_2=2.5,\mu_3=2.8$. (Left): the real part of conductivity, and (Right): the imaginary part.}  \label{n1n2n3}
\end{figure}  
   
   Accordingly, we  can define the holographic conductivity $\sigma_{\vec{\Pi}\cdot \vec{a}}$, and using the asymptotic expansion \eq{asymp} it can be expressed as
 \be\label{pcond}
 \sigma_{\vec{\Pi}\cdot \vec{a}}= -\lim_{\vec{k}\rightarrow 0} {i \sum_{j}( \rho_j a_j^{(1)} + \mu_j  a_j^{(0)})
 \over \omega \sum_{j} \mu_j a_j^{(0)} }.
\ee

   Our numerical result for $\sigma_{\vec{\Pi}\cdot \vec{a}}$ for the typical 2- and  3-band cases are shown in Fig. \ref{n1n2} and Fig. \ref{n1n2n3}, respectively. We also consider the other cases, including the 2-band one considered in the last section, and the results are similar.  We shall also mention that in our numerical calculation we find that $\sigma_{\vec{\Pi}\cdot \vec{a}}$ is independent of the choices of the initial conditions for $a_i$'s at the black hole horizon. This is not so trivial since the perturbed fields $a_i$'s are linearly coupled to each other in the bulk.  This property thus supports the identification of $\vec{\Pi}\cdot \vec{a}$ with the dual of $U(1)_{EM}$ current is sensible according to the holographic prescription. 
      
   In Fig. \ref{n1n2} and Fig. \ref{n1n2n3}, we see that  the gap appears clearly in the real part of the total conductivity Re$[\sigma_{tot}]$ as expected from the Ferrell-Glover sum rule to make up the carriers for the infinite DC conductivity. From the tail near zero frequency, we can extract the density of states for the normal component
of the carriers, i.e.,
\be n_n:= \lim_{\omega \rightarrow 0} Re[\sigma_{tot}(\omega)]\simeq
C \exp \left(-\gamma \Delta/T \right), \ee
where $\Delta:=[\sum_{i=1}^3 \langle  {\cal O}^{(i)}_2\rangle^2]^{1/4} /2$.

 Moreover, the imaginary
part of the total conductivity
 has a pole at $\omega=0$. From the Kramers-Kronig relation this implies
the infinite DC superconductivity caused by the non-zero superfluidity
 density, $n_s$.
 From our numerics, we can extract the scaling behavior
\be
n_s := \lim_{\omega \rightarrow 0} \omega Im[\sigma_{tot}
(\omega)]\simeq D (T_c-T)  \qquad \mbox{as} \quad  T\rightarrow T_c.
\ee

 The numerical fitting gives $C\sim 14$, $D\simeq 24$ and $\gamma\simeq 0.97$ for the 3-band case, this is the typical mean-field like behavior as for the holographic s-wave superconductor \cite{Hartnoll:2008vx}. For the 2-band case, we find the numerical fitting values of $C$, $D$ and $\gamma$ are not universal but depend on the chemical potentials. However, we cannot find their dependence on the chemical potentials in the closed form.

\section{Conclusion}

   In this paper, we study a holographic model which exhibits
the low energy behavior of a  multi-band superconductor.
Specifically, the two-band superconductor like $MgB_2$
has been studied quite extensively from the theoretical and as well
experimental point of view. Most of the properties of this
kind of superconducting material are believed to be explained
by the standard BCS theory, however, the insightful understanding for the appearance of the coherent order is obscure in the first principle calculation based on BCS theory. In this report we tried to construct a holographic model of this kind of multi-band superconductor, and attribute the coherent orders to the underlying emergent SO(3) symmetry.
We conjectured that the interactions among different band carriers are dictated by an underlying SO(3) global symmetry, which is then dual to the bulk SO(3) gauge dynamics. Surprisingly, our model reproduces the phase diagram with the desirable feature for multi-band superconductor, namely, the feature of the coherent orders. Moreover, we identify a gauge invariant linear combination of the perturbed gauge fields as the dual current coupled to the physical $U(1)_{EM}$ photon, and it shows the mean field BCS-like behavior for the holographic conductivity.  However, it deserves further study to clarify the physical subtlety of such an identification, especially its dependence on the background profile.  Finally, it is a natural next step to see if these features remain intact after taking into account the back reaction of the
bulk fields to the background geometry.

\bigskip \bigskip
\section*{Acknowledgements} We thank Chen-Pin Yeh for taking part
in this project at its early stage. We also thank Juinn-Wei
 Chen, Christopher Herzog and Logan Wu for helpful discussions.
This work is supported by Taiwan's NSC grant NSC-099-2811-M-003-007-, and partly by NCTS.

\section*{Appendix}

 The general definition of conductivity matrix is 
\bea
\sigma_{ij}(\omega) = \frac 1 {\omega} \lim_{\vec{k}\rightarrow 0}
\int d^4 x e^{i\omega t-i k x} \theta(t)
\langle[J_i(t,x),J_j(0,0)]\rangle.
\eea
This quantity may not be measured directly in the real experiments since it should be highly nontrivial to tune the electric field among different band carriers.  Despite that, the conductivity matrix provide some ``microscopic" picture for the holographic multi-band superconductors, and it is interesting to find out its behavior.

\begin{figure}[ht]
\center{\epsfig{figure=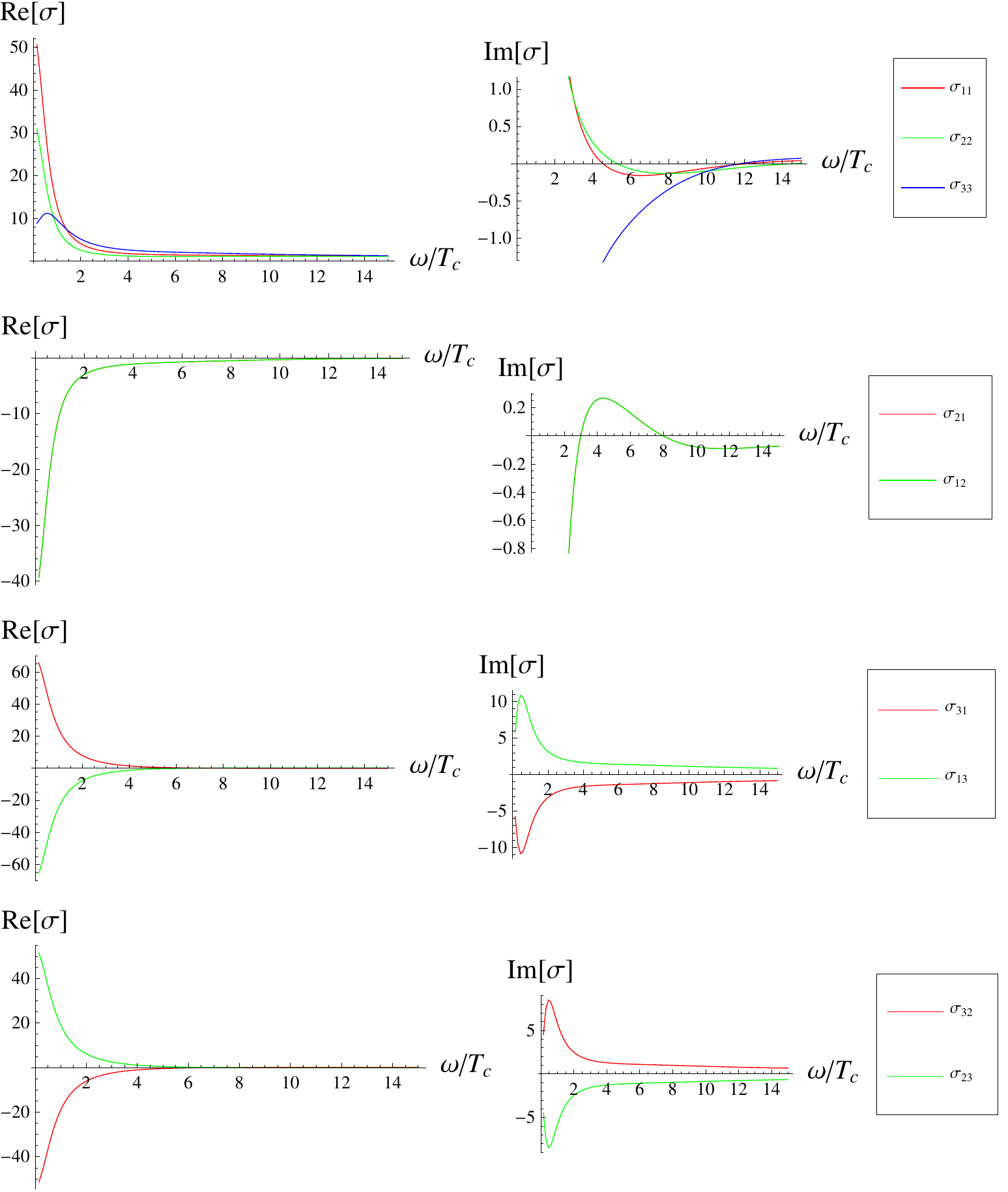,angle=0,width=16cm}}
\caption{Conductivity matrix of a 2-band superconductor for 
$\frac {T} {T_c} = 0.8460$, $\mu_2 =3.2, \mu_3=2.5$ }  \label{condmatx}
\end{figure}

  By fixing the background \eq{aansatz}, it is easy to see that 
gauge field perturbation $a_i$'s are invariant under the infinitesimal 
gauge transformation,
i.e., $\delta a_{i}=\bar{D}_x \alpha_i$ where $\bar{D}_{\mu}=
\partial_{\mu} + g \bar{A}_{\mu}$
is the covariant derivative with respect to the background
gauge field $\bar{A}_{\mu}$, and $\alpha_i$'s are the gauge parameters.
Therefore, by using AdS/CFT correspondence, 
one can define the SO(3)-invariant general conductivity
matrix as follows \footnote{This is similar to the treatment for
the holographic p-wave superconductor \cite{Gubser:2008wv}.}
 \be\label{stensor}
 \sigma_{ij}(\omega)=-\lim_{\vec{k}\rightarrow 0} {i a_i^{(1)}
(\vec{k},\omega)\over 
\omega a_j^{(0)}(\vec{k},\omega)} \mid_{a^{(0)}_{k\ne j}=0}.
\ee
A typical plot for conductivity matrix is shown in the Fig. \ref{condmatx}
for 2-band holographic superconductor. Note that it does not have the mean-field like behavior.  

\end{document}